\documentclass[superscriptaddress, twocolumn, prl]{revtex4}

\usepackage{mathrsfs, amsmath}    
\usepackage[pdftex]{graphicx}   
\usepackage{verbatim}   
\usepackage{color}      
\usepackage{subfigure}  
\usepackage{hyperref}   
\usepackage{breqn}
\usepackage{chngcntr}

\allowdisplaybreaks

\graphicspath{{figs/}}

\begin{document}
\title{A Time Domain Acoustic Model for the Production of Rodent Ultrasonic Vocalizations}

\author{Matthew Dornfeld}
\affiliation{Center for Studies in Physics and Biology\, Rockefeller University\, New York\, New York 10065}
\author{Marcelo Magnasco}
\affiliation{Center for Studies in Physics and Biology\, Rockefeller University\, New York\, New York 10065}
\author{Oreste Piro}
\affiliation{Departament de F\'isica\, Universitat de les Illes Balears\, 07122 Palma de Mallorca\, Spain}

\date{\today}
\begin{abstract}
Mammalian ultrasonic vocalization (USV) has been a subject of interest for decades. This interest has mainly been driven by the intelligence of dolphins and other odontocetes. However the semantic content of odontocete USV and its mechanism of production remain poorly understood. Serendipitously however many rodent species have convergently evolved the ability to produce USVs in a similar manner. In this paper we use rodent USV as a model process to help us gain insight into the production mechanism for mammalian USV as a whole. We derive a model that describes the production of rodent USVs by considering the interaction of an unstable jet, emerging from the vocal folds, with the passive resonance modes of the upper vocal tract. Thus our model is also a solution to a special case of the jet susceptibility problem. The derived model takes the form of a set of coupled nonlinear time domain ODEs, whose solutions are controlled by biologically relevant parameters such as subglottal pressure and vocal fold radius. In our analysis of the model we find the existence of a subglottal blowing pressure threshold ($p \approx 710$ Pa), above which steady acoustic oscillations occur. Furthermore we also reproduce the $22$ kHz rat alarm call at realistic blowing pressures ($p \approx 1500$ Pa).
\end{abstract}
\maketitle
Mammalian ultrasonic vocalization (USV) has been a subject of interest in both research and popular science literature for decades. This interest has largely been driven by the high degree of intelligence possessed by odontocetes such as dolphins and the implication that their USVs could form the building blocks of language. Despite the abundance of literature on the topic it has been very difficult for researchers to decipher the semantic meaning of the USVs or to understand the physical mechanism by which they are produced. Work on the topic is constrained by the ethical and practical limitations associated with keeping dolphins as laboratory specimens. Serendipitously however rodents have convergent evolved the ability to vocalize in the ultrasonic range in a similar manner. It seems rodents also use their USVs for communication, and we believe it is likely that the rodent USV production mechanism is similar to the one for odontocetes. Figs. \ref{fig:spectrogram_usv1} and \ref{fig:spectrogram_usv2} shows examples of spectrograms of rat USVs. This paper focuses on developing an acoustic model for the production of rodent USVs. We do this not only with the intention that such a model will help us better understand rodent USVs, but that it can also be applied to understanding mammalian USVs as a whole. Furthermore the derivation of this model will take into account the interaction between a nonlinear jet driving force and a passive acoustic resonator. This is known as the jet susceptibility problem and is an open topic of research \cite{jets2004acoustic}.

There has been a large amount of experimental and theoretical work investigating and modeling the production of sonic vocalization by vibrating vocal folds in birds and mammals. A great deal of success has been found in treating the motion of vocal folds as spring like oscillators, whose motion is driven by a pressure differential created by the lungs. However there is a good deal of evidence that mechanical oscillator models fail to explain the production of mammalian USV. Building on the work of Fletcher, who showed that for sonic vocalizations there exists a relationship of the form $f \propto M^{-0.4}$ between the fundamental vocalization frequency $f$ and the mammal's average mass $M$ \cite{Fletcher2010}, Dornfeld has shown that mammalian USVs break this scaling law by several orders of magnitude \cite{Dornfeld2017}. The divergence from the scaling law is likely due to the production mechanisms for USVs being fundamentally different from that for sonic vocalizations. Some experimental work has been done specifically investigating the mechanism of USV production in rodents. Riede and Roberts have shown that the fundamental frequencies of rodent USVs increase when the air in the vocal tract is replaced with heliox gas, a result that is incompatible with vibrating vocal folds models \cite{Riede2011} \cite{Roberts1975}. Sanders has inserted a camera into the vocal tracts of anesthetized rats and elicited vocalizations via direct brain stimulation. During vocalization he observed that the vocal folds contracted into a circular aperture with a radius of about 1 mm but did not vibrate \cite{Sanders2001}. This experimental evidence indicates the production mechanism for rodent USVs is more similar to a woodwind instrument than to vibrating vocal folds. However the underlying physics of this instrument is poorly understood. We develop a time domain acoustic model for rodent USV production by applying energy and momentum conservation to the flow of air in the rodent vocal tract.

We model the rodent upper vocal tract as a resonator driven by a jet of air emerging from the vocal folds (Fig. \ref{fig:vocal_tract}). The jet is formed when subglottal pressure $p$ induces an airflow through the vocal folds. As the jet emerges it mixes with the acoustic flow of the upper vocal tract. In addition vorticity grows on the jet boundary layer, which induces a force on the acoustic flow. The combination of these two factors drives acoustic oscillations in the upper vocal tract. The vocalization is emitted into the far field in the form of acoustic radiation through the mouth. The structure of this paper is as follows: first we will derive a time domain system that describes the temporal behavior of the upper vocal tract acoustic flow driven by a source velocity $u_0$ and vortical pressure source $p_{src}$, second we will apply energy and mass conservation to the vocal flow to derive an equation which describes the temporal behavior of $u_0$, third we will derive a time domain system that describes the behavior of the vortical jet flow and connect this flow to the acoustic flow through feedforward/feedback mechanisms, last we will perform a numerical analysis on the derived model to show that it predicts the onset of acoustic oscillations at realistic blowing pressures, and furthermore that it can also reproduce the well known $22$ kHz rat alarm call at realistic parameter values. The outline of the calculation is presented in the paper itself. However due to space restrictions the bulk of it can be found in the appendix.

To begin we will assume the acoustic flow in the upper vocal tract is described by the one dimensional potential wave equation. This is a reasonable assumption as long as wave velocity is much less than the speed of sound (an assumption which most certainly holds in the rodent vocal tract). The acoustic flow is driven by the velocity of the jet entering the upper vocal tract $u_{0}$ and is dissipated by radiation from the mouth. This forms the boundary condition at $x=0$ and $x=1$ respectively. Eq. \ref{eq:wave_appendix} shows the wave equation with these boundary conditions, in which $\phi$ is the velocity potential, and $Z$ is the radiation impedance at the mouth. The general strategy of the derivation is to move the inhomogeneous boundary condition at $x=0$ to the equation of motion itself with a substitution. The solutions $\phi_H$ of the boundary value problem with homogeneous boundary conditions are then expanded onto a set of spatial basis functions \ref{eq:appendix_phi_expansion}. The amplitudes of this expansion $q_j(t)$ govern the temporal behavior of the upper vocal tract flow and are called the modal participation factors. Inserting the expansion of $\phi_H$ into its equation of motion, integrating out the spatial variable, and algebraically solving for the modal participation factors we can show
\begin{dmath}
\label{eq:viscous}
\ddot{q}_j + \beta_j \dot{q}_j +(\omega_j^2 - \alpha_j^2) q_j = -a_j' u_0 + b_j' \ddot{u}_0,
\end{dmath} 
where $\alpha_j$ and $\omega_j$ are the real and imaginary solutions of Eq. \ref{eq:omega_alpha}. The damping constants $\beta_j$ are given by Eq. \ref{eq:beta}, and the constants $a_j'$ and $b_j'$ are given by Eq. \ref{eq:ab_constants}. The time varying quantity $u_0$ is the driving velocity entering the upper vocal tract from the vocal folds.

Now we will relate the driving velocity $u_0$ to the subglottal pressure $p$, the geometry of the vocal folds, the modal participation factors $q_j$, and the  pressure source due to vorticity $p_{src}$. We will model the flow through the vocal folds as inviscid and incompressible. The first assumption is justified by the Reynold's number of the flow ($Re \approx 2000)$. Thus it is valid to expect the flow through the vocal folds to be conservative. Therefore it is appropriate to use Bernoulli's equation to express energy conservation of the flow between a point in the trachea ($T$) and a point just outside the pharyngeal end of the vocal folds ($F$). The incompressibility assumption is justified as long as the vocal fold thickness is less than half of the shortest acoustic wavelength emitted by the vocal tract, which is on the order of $4$ mm. Although there are no precise measurements for the thickness of the rat vocal folds, since their radius is about $1$ mm it is unlikely their thickness is much larger than that, and thus it is likely the incompressibility assumption is justified. Bernoulli's equation between the points $T$ and $F$ is given by Eq. \ref{eq:energy1_appendix}. Applying mass and energy conservation to flow between the points $T$, $F$ and then enforcing pressure continuity between the point $F$ and the acoustic flow at the entrance to the upper vocal tract we get an equation that describes the temporal behavior of $u_0$.
\begin{equation}
\label{eq:u_dot}
\dot{u}_0 = \mu^{-1} \left(p - p_{src} + \sum_k{\dot{q}_k} -\frac{\gamma}{2} u_{0}^2  \right), 
\end{equation}
where the constants $\mu$ and $\gamma$ encode the geometry of the vocal folds and are given by Eq. \ref{eq:mu_gamma}, $p$ is the subglottal pressure and is a control parameter of the model. Eq. \ref{eq:u_dot} is a consistency condition, which puts an energy constraint on the solutions of Eq. \ref{eq:viscous}. It has the effect of introducing a nonlinearity into the system through its quadratic term, which is responsible for limiting the amplitude of acoustic oscillations. The nonlinear terms has the interpretation of representing the energy loss due to vortex formation as the flow passes around sharp edges \cite{Cummings1986a}.

We now take into the account the effects of the the jet detaching from the walls of the vocal folds and rolling up to form vortex rings. This component of the model is essential for the existence of sustained acoustic oscillations. A vortex ring is a tubular region of high fluid vorticity, which propagates in its axial direction. Sullivan et al. have observed the formation of vortex rings due to jet detachment from a circular orifice \cite{Sullivan2008}. In addition Chanaud and Powell have observed the formation of a street of vortex rings in the hole tone acoustic system. They also observed, in the steady state, the shedding frequency of the vortex rings is equal to the sounding frequency of the hole tone \cite{Chanaud1965}. These observations indicate vortex ring formation is most likely important in the generation of sound in acoustic systems driven by axisymmetric jets. However it is currently unclear exactly how vortex ring formation relates to the acoustic flow. This is known as the jet susceptibility problem. Here we suggest that it is the force generated by the increase in vorticity due to vortex ring formation that drives the acoustic flow. We treat this force as a point pressure source $p_{src}$ located at the origin of the upper vocal tract. With Ahkmetov's \cite{akhmetov2009vortex} expression for the force on a fluid of volume $V$ generated by a time varying vorticity $\boldsymbol{\omega}(\textbf{r},t)$ we show $p_{src}$ can be expressed by Eq. \ref{eq:p_src1_appendix}, which depends on $\frac{\partial \boldsymbol{\omega}(\textbf{r},t)}{\partial t}$. Thus to describe the dynamics of this force we need another set of equations, which govern the temporal behavior of $\boldsymbol{\omega}(\textbf{r},t)$.

To derive this dynamical system we take an approach similar to the one used to derive Eq. \ref{eq:viscous}. We will begin by assuming the axisymmetric vortex ring flow is described by the incompressible Euler equations. Vortex ring flow is historically assumed to be incompressible. The inviscid assumption is likely justified by the high Reynold's number of the flow, but it is an assumption that may need to be relaxed in future work. The axisymmetric assumption implies that the flow can be described by a Stokes stream function $\psi(r,x,t)$ defined by Eq. \ref{eq:stokes}. The $\phi$ component of the vorticity is related to the stream function by $\omega_{\phi}(r,x,t)=-\nabla^2 \psi(r,x,t)$. All other components of $\boldsymbol{\omega}$ are $0$. Thus the dynamics of $p_{src}$ can be described through this stream function. The Euler equation for $\psi$ and its boundary conditions are then given by Eqs. \ref{eq:euler_appendix} and \ref{eq:euler_bcs_appendix}. The boundary conditions at $x=0$ states that the vortical flow is driven by the velocity $u_F$ emerging from the vocal folds. Outflow boundary conditions are used at $x=1$. This is equivalent to the assumption that flow should not change much in the axial direction at $x=1$. The zero penetration boundary condition is used at $r=r_0$, and the solutions are required to remain finite at $r=0$.

We now move the inhomogeneous boundary condition at $x=0$ to the equation of motion itself with the substitution Eq. \ref{eq:psi_substitution_appendix}. We then have a PDE for $\psi_H(r,x,t)$ that has homogeneous boundary conditions and inhomogeneous driving terms in the PDE itself (Eq. \ref{eq:axi_euler2}). Out of algebraic convenience we choose the spatial basis functions $\psi_n(x)$ to be the eigenfunctions of the operator $G$. However there are most likely more efficient choices of basis functions. Future work could focus on finding such bases. The eigenvalues and eigenfunctions of $G$ are given by \ref{eq:G_solutions}. Expanding $\psi_H(r,x,t)$ onto this spatial basis $\psi_H(r,x,t)=\sum_{m=1}^{N_{\eta}} \eta_{m}(t)\psi_m(r,x)$. Then inserting it into Eq. \ref{eq:axi_euler2} and using the orthogonality relation Eq. \ref{eq:orthogonality} to integrate out the spatial basis functions, we get a dynamical system for $\eta_{n}(t)$.
\begin{dmath}
\label{eq:stream_modes}
\dot{\eta}_{n} + u_0 \sum_{m=1}^{N_{\eta}} B_{mn} \eta_{m} + \sum_{l,m=1}^{N_{\eta}}
C_{lmn} \eta_{m}\eta_{l} + u_0^2 d_{n} + \dot{u}_0 f_{n}= 0
\end{dmath} 
The spatial part of the problem is now encoded in the tensors $\textbf{B}$, $\textbf{C}$, $\textbf{d}$, and $\textbf{f}$, whose entries are given by Eq. \ref{eq:eta_spatial_constants}. Eq. \ref{eq:stream_modes} is related to the acoustic flow through the inhomogeneous driving terms proportional to $u_0^2$ and $\dot{u}_0$, as well as the convective term proportional to $u_0$.   

Now that we have a governing equation for $\boldsymbol{\eta}$ we can use that to express $p_{src}$ in terms of $u_0$, $\boldsymbol{\eta}$, and $\textbf{q}$. The result of this algebra is given by Eq. \ref{eq:p_src_appendix2}. This expression for $p_{src}$ can be inserted into Eq. \ref{eq:viscous}, which results in a differential algebraic system of equations that can be solved for $\ddot{\textbf{q}}$. The solution of this system is 
\begin{equation}
\label{eq:acoustic_oscillator}
\boldsymbol{\ddot{q}} + \boldsymbol{D} \boldsymbol{\dot{q}} + \boldsymbol{K} \boldsymbol{q} = -\textbf{a} u_0 - \textbf{b} F_b,
\end{equation}
where the tensors $\textbf{D}$, $\textbf{K}$, $\textbf{a}$, $\textbf{b}$ are given by Eqs. \ref{eq:acoustic_tensors}, and the quantity $F_b$ is given by Eq. \ref{eq:Fb_appendix}.

We have now derived the equations which govern the evolution of the acoustic flow, the driving velocity, and the driving vortical flow. Fig. \ref{fig:feedback} summarizes the relationships between these flows and the variable parameters of the model. We can integrate Eqs. \ref{eq:u_dot}, \ref{eq:stream_modes}, and \ref{eq:acoustic_oscillator}, with $p_{src}$ given by Eq. \ref{eq:p_src_appendix}, to get the temporal behavior of the system. Fig. \ref{fig:pressure_sweep_u0} shows the behavior of $u_0$ when the system is integrated from rest for different values of the subglottal pressure $p$. For this integration we choose to truncate the number of acoustic modes at $N_q=2$ because higher modes will be dissipated away by radiation, and we truncate the number of vortical modes to be $N_{\eta}=7$, because this was the lowest number of modes we found to exhibit limit cycle behavior.

\begin{figure}[!ht]
\centering
\includegraphics[width=\columnwidth]{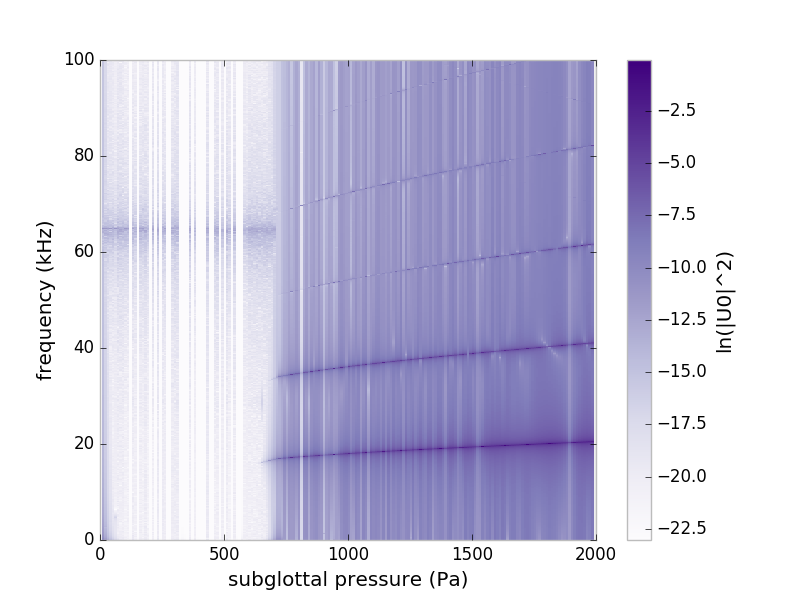}
\caption{Log scale heat map of $|U_0|^2$, the modulus squared of the Fourier transform of the driving velocity, plotted against the subglottal pressure. It can be seen that oscillations begin around $p=710$ Pa, and for $p=1500$ Pa the fundamental frequency is approximately that seen in rat alarm calls.}
\label{fig:specgram}
\end{figure}

\begin{figure}[!ht]
\centering
\includegraphics[width=\columnwidth]{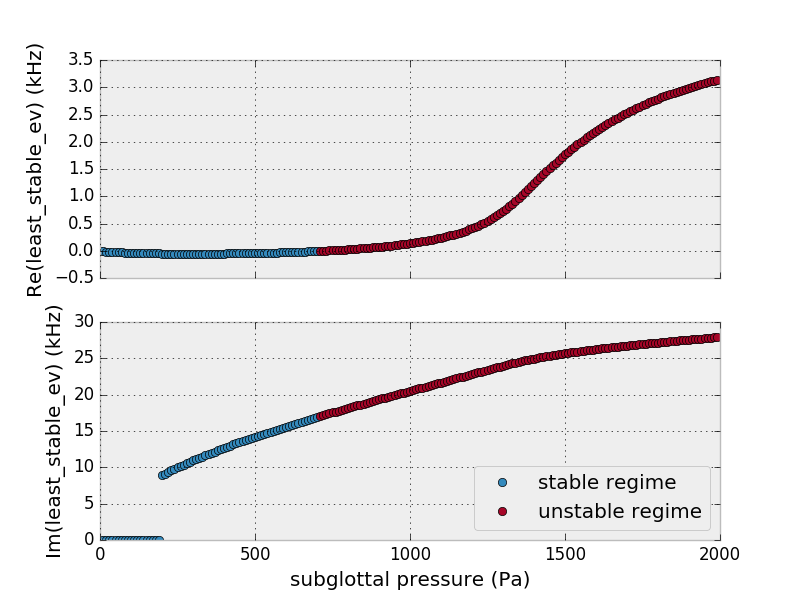}
\caption{The behavior of the dominant eigenvalue of the Jacobian evaluated at the fixed point that the system approaches when it is integrated from rest. Blue dots indicate pressures for which the real part of the least stable eigenvalue is less than $0$. Red dots indicate pressures for which the real part of the least stable eigenvalue is greater than $0$.}
\label{fig:stability}
\end{figure}

We found the subglottal pressure $p$ to be an important quantity in the control of the onset of oscillations. Fig. \ref{fig:specgram} shows the steady state frequency response of the system for different values of $p$. 
It is generated by integrating the system from rest for a range of $p$, calculating the Fourier transform $U_0(\omega) = \mathscr{F}[u_0(t)]$ (after the initial transients dies out), and plotting $\ln\left( \left| U_0 \right|^2 \right)$ as a heat map against $p$. It can be seen that for low values of $p$ oscillations will not occur, and the solutions approach a fixed point. Fig. \ref{fig:fixed_point} shows the value of $u_0$ at that fixed point. As $p$ is increased acoustic oscillations begin as the fixed point loses stability. This threshold value of $p$ can be found by calculating the eigenvalues of the Jacobian of the system, evaluated at that fixed point, and finding the value of $p$ for which the real part of the least stable eigenvalue changes signs from negative to positive. Fig. \ref{fig:stability} shows the real and imaginary parts of this least stable eigenvalue, and that the fixed point becomes unstable at $p=710$ Pa. Furthermore it can be seen in Fig. \ref{fig:specgram} that for subglottal pressures typically seen during rodent USVs ($p \approx 1500$ Pa) the fundamental frequency is approximately to 22 kHz, which is exactly the frequency of the rat alarm call! It should be noted that this result was obtained using bioligically realistic parameter values and no fitting to data. It can also be seen that overtones are in the same range as the higher frequency rat social calls ($60-85$ kHz). However due to the complexity of the parameter space and the importance of attack transients in the determination of the steady state we had trouble finding parameter values that resulted in oscillations with the energy concentrated in this frequency range. This could be a future area of exploration.

To summarize what has been done: a dynamical system that describes the transient behavior of acoustic oscillations in the rodent vocal tract has been derived. It has been shown that acoustic oscillations begin with a loss of stability of a fixed point as the subglottal pressure is increased. Once the oscillation threshold has been passed further increasing the subglottal pressure causes modulation of the emitted acoustic frequency. These are all features that have been observed experimentally in rodent USVs. However one notable property of rodent USVs, the discontinuous jumping of vocalization frequency from one value to another, has not been recreated using this model. It is difficult to explore the full parameter space of the model, and determining its full bifurcation structure will require a more sophisticated analysis. Thus it is possible frequency jumps occur in some regime yet explored. It is also possible a mechanism has been left out of this model. It is our belief that considering the inertia of the jet as it is absorbed into the upper vocal tract will introduce a time delay factor into the time domain dynamical system governing the acoustic oscillations. We believe this may be an important mechanism, which will produce frequency jumps in the model, since previous modeling of acoustic systems with time delays have produced models with frequency jumps \cite{Arneodo2009}, \cite{Auvray2012a}.

\appendix 
\newcounter{eqncounter}
\setcounter{eqncounter}{1}
\renewcommand\theequation{A.\theeqncounter}
\newcounter{figcounter}
\setcounter{figcounter}{1}
\renewcommand\thefigure{A.\thefigcounter}

\section{Appendix}
\subsection{Spectrogram of Rodent Ultrasonic Vocalization}
\begin{figure}[!ht]
\centering
\includegraphics[width=\columnwidth]{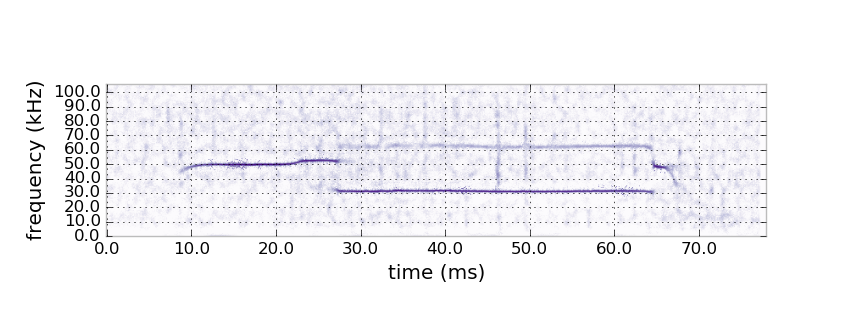}
\caption{Spectrogram of a rat USV. This particular call snippet exhibits two frequency jumps.}
\label{fig:spectrogram_usv1}
\end{figure}
\stepcounter{figcounter}

\begin{figure}[!ht]
\centering
\includegraphics[width=\columnwidth]{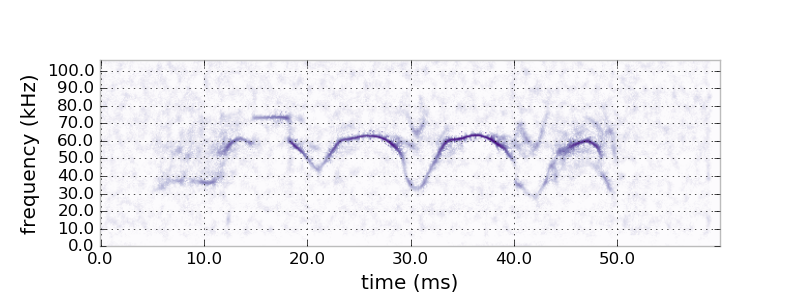}
\caption{Spectrogram of a rat USV. This call snippet is towards the upper frequency limit of the rat's ability to vocalized. These calls tend to show more frequency modulation than lower frequency calls and are associated with positive mental states.}
\label{fig:spectrogram_usv2}
\end{figure}
\stepcounter{figcounter}

\subsection{The Upper Vocal Tract Flow}
\begin{figure}[!ht]
\centering
\includegraphics[width=\columnwidth]{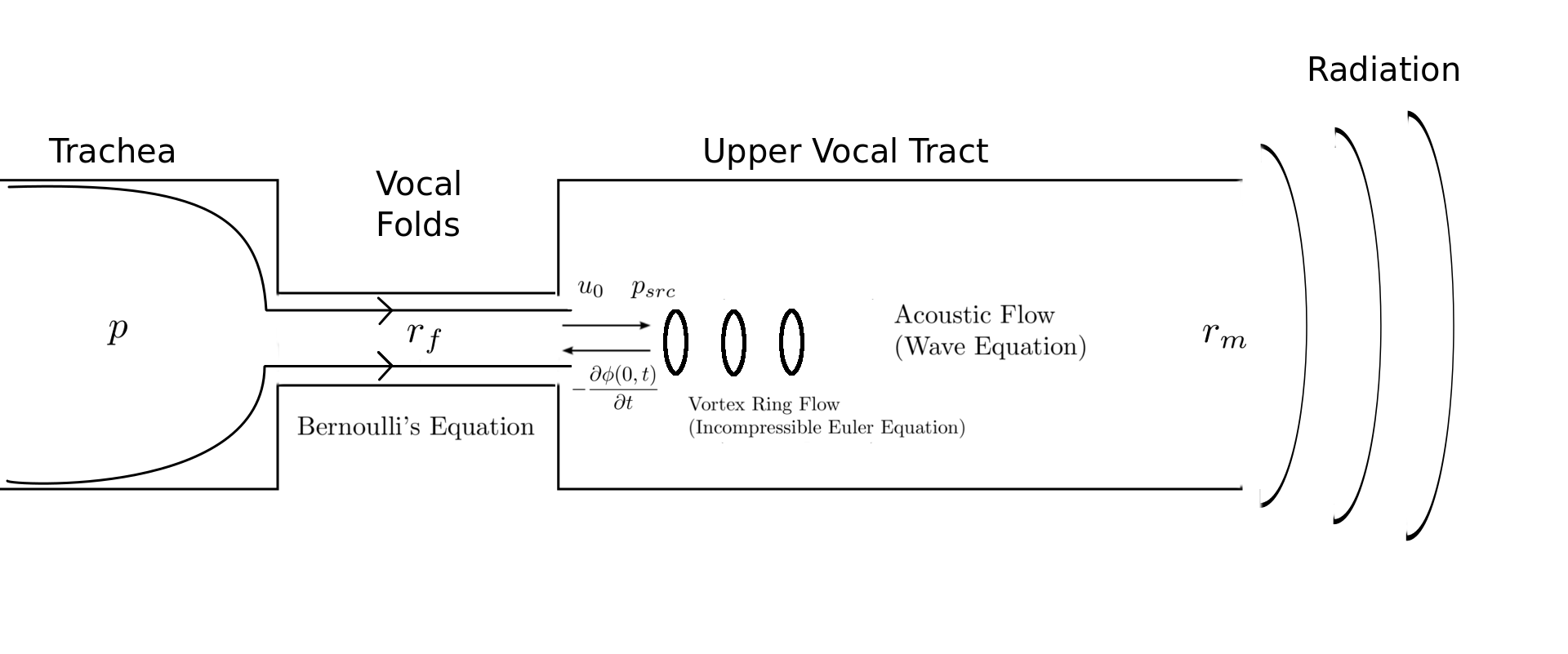}
\caption{A schematic model of the rodent vocal tract. Subglottal pressure $p$ drives flow through the vocal folds, which have radius $r_f$. The flow $u_0$, emerging from the vocal folds drives the acoustic flow in the upper vocal tract, which is described by the velocity potential $\phi(x,t)$. The acoustic flow is damped by radiation at the mouth, which is controlled by the mouth radius $r_m$. In addition the jet detaching from the walls of the vocal folds leads to the formation of vortex rings described by the stream function $\psi$, which leads to an additional pressure $p_{src}$ on the acoustic flow. The acoustic flow provides feedback pressure $-\frac{\partial \phi(0,t)}{\partial t}$, which opposes $u_0$.}
\label{fig:vocal_tract}
\end{figure}

Fig. \ref{fig:vocal_tract} shows a schematic of the model presented in this paper. To begin we will assume the acoustic flow in the upper vocal tract is described by the one dimensional potential wave equation. This is a reasonable assumption as long as wave velocity is much less than the speed of sound (an assumption which most certainly holds in the rodent vocal tract). The acoustic flow is driven by the velocity of the jet entering the upper vocal tract $u_{0}$ . This forms the boundary condition at $x=0$. The acoustic flow is dissipated by radiation from the mouth. This forms the boundary condition at $x=1$
\begin{equation}
\label{eq:wave_appendix}
\begin{split}
\frac{\partial^2 \phi}{\partial x^2} &- \frac{\partial^2 \phi}{\partial t^2} = 0 \\ 
\frac{\partial \phi (0,t)}{\partial x} = u_0 (t) &\quad
\frac{\partial \phi (1,t)}{\partial t} + Z \frac{\partial \phi (1,t)}{\partial x} = 0 \\
\end{split}
\end{equation}
\stepcounter{eqncounter}
Here $\phi$ is the velocity potential, and $Z$ is the radiation impedance at the mouth (this will be discussed more later). The driving velocity $u_0$ is allowed to depend on the velocity potential and its time derivative through feedback.
\stepcounter{figcounter}

To perform the spatial eigenfunction expansion the inhomogeneous boundary condition at $x=0$ must be moved into the equation of motion itself. This can be done with the substitution $\phi(x,t)=\phi_H (x,t)-u_0 \frac{(x-1)^2}{2}$, where $\phi_H (x,t)$ is a function that satisfies the homogeneous boundary conditions but has an inhomogeneous driving term in the wave equation.
\begin{equation}
\label{eq:driven_wave}
\begin{split}
\frac{\partial^2 \phi_H}{\partial x^2} & - \frac{\partial^2 \phi_H}{\partial t^2} = u_0 - \ddot{u}_0 \frac{(x-1)^2}{2} \\
\frac{\partial \phi_H (0,t)}{\partial x} = 0 & \quad  
\frac{\partial \phi_H (1,t)}{\partial t} + Z \frac{\partial \phi_H (1,t)}{\partial x} = 0 \\
\end{split}
\end{equation}\stepcounter{eqncounter} 
Next expanding $\phi_H (x,t)$ onto a set of spatial basis functions $\phi_j (x)$.
\begin{equation}
\label{eq:appendix_phi_expansion}
\phi_H (x,t) = \sum_{m=1}^\infty q_j(t) \phi_j (x)
\end{equation}
The projections $q_j(t)$ vary with time and are called the modal participation factors. The spatial basis functions $\phi_j (x)$ must be chosen in such a way that the solutions satisfy the boundary conditions in Eq. \ref{eq:driven_wave}
\begin{equation}
\label{eq:coupled_equation}
\sum_{m=1}^\infty q_j \phi_j''(x) - \ddot{q}_j \phi_j(x) = u_0 - \ddot{u}_0 \frac{(x-1)^2}{2}.
\end{equation}\stepcounter{eqncounter}

\begin{figure}
\centering
\includegraphics[width=\linewidth]{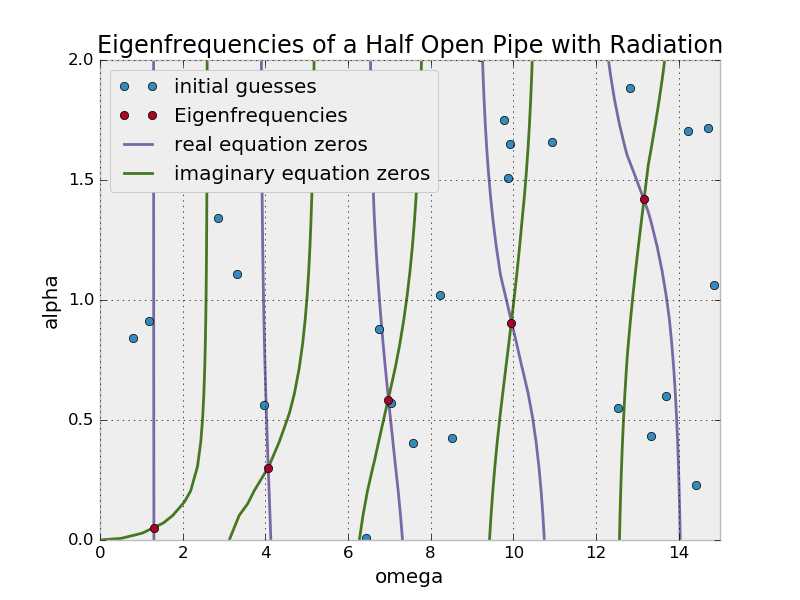}
\caption{Solutions of Eq. \ref{eq:omega_alpha} for $r_m=4 \times 10^{-6}$ ($1$ mm in dimensional units). The purple and green curves show the zero isoclines for the real and imaginary parts of Eq. \ref{eq:omega_alpha}. The intersections of these curves are the eigenfrequencies of the half open pipe. The fsolve function, which is a part of the scipy.optimize package was used to solve this system. The blue circles show initial guesses supplied to fsolve. The red circles show the output of fsolve. These are the eigenfrequencies of the system.}
\label{fig:eigenfrequencies}
\end{figure}

We are free to choose the spatial basis $\phi_j(x)$ however we want as long as the solutions satisfy the boundary conditions in Eq. \ref{eq:driven_wave}. However we can simplify the calculation by choosing the basis functions to be the solutions of the boundary value problem
\begin{equation}
\label{eq:spatial_problem}
\begin{split}
\phi_j''(x)-s_j^2 \phi_j(x) = 0 \\
\phi_j'(0)=0 \\
s_j \phi_j(1) + \hat{Z}(\omega_j) \phi_j'(1)=0
\end{split}
\end{equation}\stepcounter{eqncounter}
Here $s_j=-\alpha_j + i \omega_j$ and $\hat{Z}(\omega)$ is the Fourier transform of $Z$, the radiation impedance at the mouth. The effects of radiation can be taken into account by modeling the end of the vocal tract at $x=1$ as a moving piston that radiates spherical sound waves off into the far field. Thinking of the far end of the vocal tract this way the Fourier transform of the radiation impedance can be written as
\begin{equation}
\label{eq:impedance}
\begin{split}
\hat{Z}(\omega) &= R(\omega) + i X(\omega) \\
&=\left(1 - \frac{J_1(2 \omega r_m)}{\omega r_m} + i \frac{H_1(2 \omega r_m)}{\omega r_m} \right).
\end{split}
\end{equation}\stepcounter{eqncounter}
Here $r_m$ is the radius of the rodent mouth, $J_1$ is the Bessel function of the first kind of order 1, and $H_1$ is the Struve function of order 1 \cite{pierce1981acoustics}. After applying the boundary condition at $x=0$ the solution to Eq. \ref{eq:spatial_problem} becomes
\begin{equation}
\label{eq:spatial_basis}
\phi_j(x) = \cosh(s_j x).
\end{equation}\stepcounter{eqncounter}
Substituting Eqs. \ref{eq:spatial_basis} and \ref{eq:impedance} into the boundary condition at $x=1$ a complex equation is obtained for $\omega_j$ and $\alpha_j$. 
\begin{equation}
\begin{split}
\label{eq:omega_alpha}
\begin{pmatrix}
\cos(\omega) \sinh(\alpha) & \sin(\omega) \cosh(\alpha) \\
\sin(\omega) \cosh(\alpha) & -\cos(\omega) \sinh(\alpha) 
\end{pmatrix}
\begin{pmatrix}
R(\omega) \\
X(\omega)
\end{pmatrix}= \\
\begin{pmatrix}
\cos(\omega) \cosh(\alpha) \\
\sin(\omega) \sinh(\alpha)
\end{pmatrix}
\end{split}
\end{equation}\stepcounter{eqncounter}
This equation can be solved numerically to obtain these quantities. Fig. \ref{fig:eigenfrequencies} shows an example of this numerical calculation.
\stepcounter{figcounter}

From Eq. \ref{eq:spatial_basis} it can be seen that $\phi_j''(x) = - s_j^2 \phi_j(x)$. Substituting this relation into Eq. \ref{eq:driven_wave},
\begin{equation}
\label{eq:coupled_equation2}
\sum_{m=1}^\infty (q_j s_j^2 - \ddot{q}_j) \phi_j(x) = u_0 - \ddot{u}_0 \frac{(x-1)^2}{2}.
\end{equation}\stepcounter{eqncounter}
This can be written as a matrix equation by defining the vector $v_j = q_j s_j^2 - \ddot{q}_j$ and matrix $ \boldsymbol{\Phi}_{xm} = \phi_j(x)$.
\begin{equation}
\label{eq:matrix_eq1}
\boldsymbol{\Phi}.\boldsymbol{v} = \boldsymbol{1} u_0 - \frac{(\boldsymbol{x}-1)^2}{2} \ddot{u}_0,
\end{equation}\stepcounter{eqncounter}
where $\boldsymbol{1}$ is the vector with all ones as entries and $\boldsymbol{x}$ is the position vector. The goal now it solve the differential algebraic equation \ref{eq:matrix_eq1} for a set of ordinary differential equations. Because of the inclusion of radiation in the right boundary condition the rows of $\boldsymbol{\Phi}$ are not orthogonal to each other. They can be rotated into a basis in which they are orthogonal by diagonalizing the overlap matrix $S_{ij} = \int_{0}^{1} \phi_i^*(x) \phi_j(x) dx$. This can be done numerically by computing the eigendecomposition of the overlap matrix $\boldsymbol{S}=\boldsymbol{T} \boldsymbol{\Lambda} \boldsymbol{T}^\dagger$. The matrix $\boldsymbol{T}$ possesses the eigenvectors of $\boldsymbol{S}$ as columns and is unitary. The matrix $\boldsymbol{\Lambda}$ is diagonal and has the eigenvalues of $\boldsymbol{S}$ as its nonzero entries. Hence $\boldsymbol{T}$ rotates $\boldsymbol{S}$ into a basis in which it is diagonal and thus a basis in which the rows of $\boldsymbol{\Phi}$ are orthogonal. We write this new matrix with orthogonal rows as $\widetilde{\boldsymbol{\Phi}}=\boldsymbol{\Phi}.\boldsymbol{T}$. Since $\boldsymbol{T}$ is unitary Eq. \ref{eq:matrix_eq1} can be written as
\begin{equation}
\label{eq:matrix_eq2}
\boldsymbol{\Phi}.\boldsymbol{T}.\boldsymbol{T^\dagger}.\textbf{v} = \textbf{1} u_0 - \frac{(\textbf{x}-1)^2}{2} \ddot{u}_0
\end{equation}\stepcounter{eqncounter}
Thus,
\begin{equation}
\label{eq:matrix_eq3}
\widetilde{\boldsymbol{\Phi}}.\boldsymbol{T^\dagger}.\textbf{v} = \textbf{1} u_0 - \frac{(\textbf{x}-1)^2}{2} \ddot{u}_0
\end{equation}\stepcounter{eqncounter}
Now multiplying this equation by $\widetilde{\boldsymbol{\Phi}}^\dagger$,
\begin{equation}
\label{eq:matrix_eq4}
\boldsymbol{\Lambda}.\boldsymbol{T^\dagger}.\textbf{v} = \widetilde{\boldsymbol{\Phi}}^\dagger. \left( \textbf{1} u_0 - \frac{(\textbf{x}-1)^2}{2} \ddot{u}_0 \right),
\end{equation}\stepcounter{eqncounter}
using the fact that $\widetilde{\boldsymbol{\Phi}}^\dagger.\widetilde{\boldsymbol{\Phi}}=\boldsymbol{\Lambda}$. Now multiplying this equation by $(\boldsymbol{\Lambda}.\boldsymbol{T^\dagger})^{-1}$,
\begin{equation}
\label{eq:matrix_eq4}
\textbf{v} = \boldsymbol{T^\dagger}.\boldsymbol{\Lambda}^{-1}.\widetilde{\boldsymbol{\Phi}}^\dagger. \left( \textbf{1} u_0 - \frac{(\textbf{x}-1)^2}{2} \ddot{u}_0 \right).
\end{equation}\stepcounter{eqncounter}
Converting this equation back to component form it becomes
\begin{dmath}
\label{eq:hysteretic_damping_appendix}
\ddot{q}_j + (\omega_j^2 - \alpha_j^2) \left( 1 +  i \frac{2 \alpha_j \omega_j}{\omega_j^2 - \alpha_j^2} \right) q_j = -a_j' u_0 + b_j' \ddot{u}_0, 
\end{dmath}\stepcounter{eqncounter}
where 
\begin{equation}
\label{eq:ab_constants}
\begin{split}
a_j' &= (\boldsymbol{T}.\boldsymbol{\Lambda}^{-1}.\widetilde{\boldsymbol{\Phi}}^\dagger.\boldsymbol{1})_j \\
b_j' &= (\boldsymbol{T}.\boldsymbol{\Lambda}^{-1}.\widetilde{\boldsymbol{\Phi}}^\dagger.\frac{(\boldsymbol{x}-1)^2}{2})_j
\end{split}
\end{equation}\stepcounter{eqncounter}

Eq. \ref{eq:hysteretic_damping_appendix} describes a set of driven damped oscillator equations. The terms proportional to $i q_j$ are known as hysteretic damping terms. They are similar to viscous damping terms in that the multiplication by $i$ makes them $\frac{\pi}{2}$ radians out of phase with the elastic and inertial terms. However the hysteretic damping term presents a problem when it is including in time domain simulations, namely it admits acausal solutions, which cause numerical simulations to diverge. The interpretation of Eq. \ref{eq:hysteretic_damping_appendix} is that only the real part of the submanifold of the solution space, which does not violate causality, has physical meaning. Since the solutions of Eq. \ref{eq:hysteretic_damping_appendix} are complex this gives us twice the additional degrees of freedom in choosing the initial conditions. In theory we can choose the initial conditions to ensure the solutions remain on the causal submanifold, ensuring the solutions are physically meaningful. In practice this is difficult to do and the presence of numerical noise can perturb solutions from the physically meaningful submanifold resulting in the acausal part of the solutions to grow exponentially. A more tractable way of dealing with the hysteretic damping term is to approximate it by a viscous damping term. The idea is then to choose the coefficients of the viscous damping terms so that the solutions approximate the causal solutions of the hysteretic damping equation equation. This is similar to the procedure discussed by Henwood in \cite{Henwood2002}. More concretely, if we have an equation of the form
\begin{dmath}
\label{eq:viscous_approximation_appendix}
\ddot{q}_j + \beta_j \dot{q}_j +(\omega_j^2 - \alpha_j^2) q_j = -a_j' u_0 + b_j' \ddot{u}_0,
\end{dmath}\stepcounter{eqncounter} 
can we find $\beta_j$ such that the solutions of Eq. \ref{eq:viscous_approximation_appendix} approximate the solutions of Eq. \ref{eq:hysteretic_damping_appendix}. To do this we denote the natural frequencies of the hysteretic and viscous damping equations as $\omega_h$ and $\omega_v$. The idea is then to minimize the square of the distance between these frequencies with respect to $\beta_j$. In essence this amounts to solving the equation $\frac{\partial}{\partial \beta_j} |\omega_h - \omega_v|^2 = 0$, which is satisfied when 
\begin{equation}
\label{eq:beta}
\beta_j = 2 \sqrt{\omega_j^2 - \alpha_j^2} \sin \left( \frac{1}{2}\tan^{-1} \left( \frac{2 \alpha_j \omega_j}{\omega_j^2 - \alpha_j^2} \right) \right).
\end{equation}\stepcounter{eqncounter}
Thus if the viscous damping coefficients are given by Eq. \ref{eq:beta}, the oscillations in the upper vocal tract can be approximately described by the solutions of Eq. \ref{eq:viscous_approximation_appendix}, a set of damped harmonic oscillator equations driven by the velocity entering the upper vocal tract $u_0$.

\subsection{The Flow Through the Vocal Folds}
Bernoulli's equation between the points $T$ and $G$ is given by 
\begin{equation}
\label{eq:energy1_appendix}
(p_{G} - p_{T}) + \frac{1}{2}(u_{G}^2 - u_{T}^2) + \frac{\partial}{\partial t} (\phi_{G} - \phi_{T}) = 0.
\end{equation}\stepcounter{eqncounter}  
Here $p_{G,T}$ are the pressures at those points, $u_{G,T}$ are the velocities at those points, and $\phi_{G,T}$ are the velocity potentials at those points. From continuity of mass and the incompressibility of the flow, $u_{T}=\frac{A_{G}}{A_{T}}u_{G}$. Furthermore, the change in velocity potential can be expressed as $\phi_{G} - \phi_{T} = \int_{T}^{G} u(x) dx$, where $u(x)$ is the axially varying flow velocity inside the folds. Using conservation of mass again this can be written in terms of the velocity at the glottal end $\phi_{G} - \phi_{T} = A_{G} u_{G} \int_{T}^{G} \frac{dx}{A_F}=l u_F$, where $l$ is the length of the vocal folds. Inserting these equations into Eq. \ref{eq:energy1_appendix}  
\begin{equation}
\label{eq:energy2}
(p_{G} - p_{T}) + \frac{1}{2} u_{G}^2 \left(1 - \left( \frac{A_{G}}{A_{T}} \right)^2 \right) + \frac{\partial}{\partial t} \left( u_{G} l \right) = 0.
\end{equation}\stepcounter{eqncounter} 
The velocity $u_{G}$ will be fed into the upper vocal tract exciting acoustic resonances. This is the feed-forward element of the system. 
marker
After a short distance the flow emerging from the vocal folds will mix with the acoustic flow of the upper vocal tract. Conservation of mass can be used again to relate the flow emerging from the vocal folds to $u_0$, the driving velocity in Eq. \ref{eq:viscous_approximation_appendix}, $A_{G} u_{G} = u_0 A_0$. Inserting this into Eq. \ref{eq:energy2},
\begin{dmath}
\label{eq:energy3}
(p_{G} - p_{T}) + \frac{1}{2} u_{0}^2 \left( \left( \frac{A_{0}}{A_{G}} \right)^2- \left( \frac{A_{0}}{A_{T}} \right)^2 \right) + \frac{\partial}{\partial t} \left( \frac{A_{0}}{A_F} u_{0} l \right) = 0.
\end{dmath}\stepcounter{eqncounter} 
By continuity of pressure $p_{G}$ must equal the acoustic pressure response generated in the upper vocal tract at $x=0$ plus the pressure from any additional sources driving the acoustic flow. Thus, $p_{G}=p_{src} + p(0,t)= p_{src} -\frac{\partial \phi(0,t)}{\partial t}= p_{src} -\sum_l{\dot{q}_k} + \frac{\dot{u}_0}{2}$. This is the feedback condition, since the driving velocity $u_0$ will excite a pressure response in the resonator, which will impede further flow. The nature of $p_{src}$ will be discussed more later. Again by continuity of pressure $p_{T}$ must equal the pressure in the trachea. This is the input to the system, and it will just be called $p$, with the understanding that it can be made to vary in time. Inserting these equations into Eq. \ref{eq:energy3}, 
\begin{dmath}
\label{eq:energy4}
p_{src} -\sum_l{\dot{q}_k} + \frac{\dot{u}_0}{2} - p + \frac{1}{2} u_{0}^2 \left( \left( \frac{A_{0}}{A_{G}} \right)^2- \left( \frac{A_{0}}{A_{T}} \right)^2 \right) + \frac{\partial}{\partial t} \left( \frac{A_{0}}{A_F} u_{0} l \right) = 0.
\end{dmath}
This equation can be simplified by defining the area ratios $C_E = \frac{A_{0}}{A_{G}}$ (expansion coefficient) and $C_C = \frac{A_F}{A_t}$ (contraction coefficient).
\begin{dmath}
\label{eq:energy5}
p_{src}-\sum_l{\dot{q}_k} + \frac{\dot{u}_0}{2} - p + \frac{C_E^2}{2} u_{0}^2 \left( 1 - C_C^2 \right) + \frac{\partial}{\partial t} \left( C_E u_{0} l \right) = 0.
\end{dmath}

Is is also reasonable to assume the time derivatives of $C_E$ and $l$ are negligible compared to those of $u_0$ and $q_n$, since the acoustic oscillations have a much higher frequency than is physically possible for the vocal folds to maintain. Thus we can take them out of the time derivative in Eq. \ref{eq:energy5}
\begin{equation}
\label{eq:energy7}
p_{src}-\sum_l{\dot{q}_k} - p + \frac{1}{2} u_{0}^2 C_E^2 \left( 1 - C_C^2 \right) + \left( C_E l + \frac{1}{2}  \right) \dot{u}_0 = 0.
\end{equation}\stepcounter{eqncounter} 
Solving this for $\dot{u}_0$ and gathering the constants together we get
\begin{equation}
\label{eq:u_dot_appendix}
\dot{u}_0 = \mu^{-1} \left(p - p_{src} + \sum_l{\dot{q}_k} -\frac{\gamma}{2} u_{0}^2  \right),
\end{equation}\stepcounter{eqncounter}
where
\begin{equation}
\label{eq:mu_gamma}
\begin{split}
\mu &= C_E l + \frac{1}{2} = \frac{r_0}{r_f}l + \frac{1}{2} \\ 
\gamma &= C_E^2 \left( 1 - C_C^2 \right) = \left( \frac{r_0}{r_f} \right)^2 \left( 1 - \left( \frac{r_f}{r_t} \right)^2 \right)
\end{split}
\end{equation}\stepcounter{eqncounter}
\ref{eq:u_dot_appendix} is a consistency condition, which puts an energy constraint on the solutions of Eq. \ref{eq:viscous_approximation_appendix}. It has the effect of introducing a nonlinearity into the system through its quadratic term, which is responsible for limiting the amplitude of acoustic oscillations. The nonlinear terms has the interpretation of representing the energy loss due to vortex formation as the flow passes around sharp edges \cite{Cummings1986a}.

\subsection{The Unsteady Pressure Source Due to Vortex Ring Formation}
In the derivations of the previous sections we neglected an important mechanism, the jet detaching from the walls of the vocal folds and rolling up to form vortex rings. A vortex ring is a tubular region of high vorticity in a fluid, which propagates in its axial direction. Sullivan et al. have observed the formation of vortex rings due to jet detachment from a circular orifice \cite{Sullivan2008}. In addition Chanaud and Powell have observed the formation of a street of vortex rings in the hole tone acoustic system. They also observed, that in the steady state, the shedding frequency of the vortex rings is equal to the sounding frequency of the hole tone \cite{Chanaud1965}. These observations indicate vortex ring formation is most likely important in the generation of sound in acoustic systems driven by axisymmetric jets. However it is currently unclear exactly how vortex ring formation relates to the acoustic flow. Here we suggest that the force generated by the increase in vorticity due to vortex ring formation drives the acoustic flow. We will treat this force as a point pressure source at the origin of the upper vocal tract $p_{src}$. We will also derive a system of equations that govern the temporal evolution of the vortex ring flow and show the presence of a feedback mechanism, which modulates this flow through the velocity $u_0$.

Ahkmetov has shown that for a time varying vorticity $\boldsymbol{\omega}(\textbf{r},t)$ in a fluid of volume $V$ the force on the fluid is given by \cite{akhmetov2009vortex}
\begin{equation}
\label{eq:F_omega}
\textbf{F}_{\omega} = \frac{1}{2} \int_{V} dV \left( \textbf{r} \times \frac{\partial \boldsymbol{\omega}(\textbf{r},t)}{\partial t} \right)
\end{equation}\stepcounter{eqncounter}
In the previous sections, because of the axial symmetry and the high cut on frequency of the radial modes, we restricted the acoustic flow to the axial dimension of the upper vocal tract. Thus only the axial component of the force in Eq. \ref{eq:F_omega} is relevant. We can treat that force as one a dimensional pressure source at the origin of the upper of the vocal tract. Thus we can write
\begin{equation}
\label{eq:p_src1_appendix}
p_{src} = \frac{(\textbf{F}_{\omega})_x}{\pi r_0^2} =  \frac{1}{2 \pi r_0^2} \int_{0}^{r_0}\int_{0}^{1} r^2 \frac{\partial \omega_{\phi}(r,x,t)}{\partial t} dr dx
\end{equation}\stepcounter{eqncounter}
If the flow is incompressible and axisymmetric with axial and radial velocities given by $u$ and $v$. The flow can be described by a Stokes stream function defined by
\begin{equation}
\label{eq:stokes}
u = \frac{1}{r} \frac{\partial \psi}{\partial r} \quad v = - \frac{1}{r} \frac{\partial \psi}{\partial x}.
\end{equation}\stepcounter{eqncounter}
The azimuthal component of the vorticity can then be expressed as $\omega_{\phi}(r,x,t)=-\nabla^2 \psi(r,x,t)$. Thus the pressure source can be expressed as
\begin{equation}
\label{eq:p_src_appendix}
p_{src} = \frac{(\textbf{F}_{\omega})_x}{\pi r_0^2} =  -\frac{1}{2 \pi r_0^2} \int_{0}^{r_0}\int_{0}^{1} r^2 \frac{\partial}{\partial t}\nabla^2 \psi(r,x,t) dr dx.
\end{equation}\stepcounter{eqncounter}
With Eq. \ref{eq:p_src_appendix} we can include the forcing due to vortex ring formation on Eqs. \ref{eq:u_dot_appendix} and \ref{eq:viscous_approximation_appendix}. However to fully describe the dynamics of the system we need another set of ODEs that describe the time domain behavior of $\psi(r,x,t)$. We will derive these equations by expanding $\psi$ onto a set of spatial basis functions, integrating out the spatial functions, which will leave us with a set of time domain equations that describe the evolution of the coefficients in the expansion.

Beginning with the incompressible Euler equation in stream function form,
\begin{equation}
\label{eq:euler_appendix}
\frac{\partial }{\partial t} G \psi + \frac{1}{r}\left( D_r \psi D_x - D_x \psi H\right) G \psi = 0
\end{equation}\stepcounter{eqncounter}
where $G= D_r^2 - \frac{D_r}{r} + D_x^2$ and $H=D_r - \frac{2}{r}$. Here the derivatives with respect to $x$ and $r$ are written as $D_x$ and $D_r$. The boundary conditions are
\begin{equation}
\label{eq:euler_bcs_appendix}
\begin{split}
D_r \psi(r,0,t) = u_F(t) r P(r) &  \quad  D_r D_x \psi(r,1,t)=0 \\
\left| \frac{1}{r}D_r \psi(0, x, t) \right| < \infty & \quad D_x \psi(r_0, x, t) =0
\end{split} 
\end{equation}\stepcounter{eqncounter}
Here $P(r)$ is the hyperbolic tangent jet profile given by
\begin{equation}
\label{eq:velocity_profile}
P(r)=\frac{1}{2} \left(1 + \tanh \left(\frac{1}{4 \theta} \left(\frac{r_f}{r} - \frac{r}{r_f} \right) \right) \right),
\end{equation}\stepcounter{eqncounter}
where $\theta$ is the momentum thickness of the jet. It takes into the account that the velocity at the vocal folds is $u_F(t)$ for $r<r_f$ and approximately $0$ outside that radius.

To perform the analysis, first the inhomogeneity in the boundary conditions must be transferred to the equation itself. This can be done with the substitution 
\begin{equation}
\begin{split}
\label{eq:psi_substitution_appendix}
\psi(r,x,t) &= \psi_H(r,x,t)+u_0(t) \chi_0(r,x) \\
\chi_0(r,x) &= C_E (x-1)^2 \int r P(r) dr.
\end{split}
\end{equation}\stepcounter{eqncounter}
Here $u_{0}(t)$ is the velocity entering the upper vocal tract. This quantity will be important in connecting the vortical flow back to the acoustic flow. With this substitution the equation of motion becomes
\begin{equation}
\label{eq:axi_euler2}
\begin{split}
& \left( D_t + \frac{1}{r} ( D_r u_0(t) \chi_0 D_x - D_x u_0(t) \chi_0 H) \right) G \psi_H \\
&+ \frac{1}{r}(D_x G u_0(t) \chi_0 D_r - H G u_0(t) \chi_0 D_x ) \psi_H \\ 
&+ \frac{1}{r}\left( D_r \psi_H D_x G \psi_H - D_x \psi_H H G \psi_H\right) \\
&+ \frac{1}{r}\left( D_r u_0(t) \chi_0 D_x G u_0(t) \chi_0 - D_x u_0(t) \chi_0 H G u_0(t) \chi_0 \right) \\ 
& + \dot{u}_0(t) G\chi_0 = 0 
\end{split}
\end{equation}\stepcounter{eqncounter}
The boundary conditions become.
\begin{equation}
\begin{split}
D_r \psi_H(r,0,t) = 0 &  \quad  D_r D_x \psi_H(r,1,t)=0 \\
\left| \frac{1}{r}D_r \psi_H(0, x, t) \right| < \infty & \quad D_x \psi_H(r_0, x, t) =0
\end{split} 
\end{equation}\stepcounter{eqncounter}
Because of it's position in Eq. \ref{eq:axi_euler2} the solutions of the eigenvalue problem associated with the operator $G$ are a convenient choice for the required spatial eigenfunctions. The eigenfunctions can be obtained by solving the Sturm-Liouville boundary value problem.
\begin{equation}
\begin{split}
\label{eq:G_eig}
& G \psi_{ij}(r,x) = \lambda_{ij} \psi_{ij}(r,x) \\
& \psi_{ij}(r,0,t) = 0 \quad \frac{\partial \psi_{ij}(r,1,t)}{\partial x}=0 \\ 
& \left| \frac{1}{r}D_r\psi_{ij}(0, x, t) \right| < \infty \quad \psi_{ij}(r_0, x, t)=0 \quad  
\end{split}
\end{equation}\stepcounter{eqncounter}
It should be noted that the eigenfunctions of $G$ are chosen as a basis because of algebraic convenience. There most likely exists a basis, which more efficiently captures the dynamics of the solutions of Eq. \ref{eq:axi_euler2}. The solutions of the eigenvalue problem are given by
\begin{equation}
\begin{split}
\psi_{ij}(r,x) &= r J_1 (j_{1,i} \frac{r}{r_0}) \sin \left( \left(j-\frac{1}{2} \right) \pi x\right) \\ \lambda_{ij} &= \left( \frac{j_{1,i}}{r_0} \right)^2 + (2j-1)^2 \frac{\pi^2}{4} \quad i,j \geq 1
\end{split}
\end{equation}\stepcounter{eqncounter}
The eigenfunctions are indexed by two subscripts, one for the axial modes and one for the radial ones. For notational convenience these two subscripts will be combined together with the following transformation. If the number of modes are truncated to $N_x$ axial modes and $N_r$ radial modes. The subscripts $i$ and $j$ can be expressed as a function of a single subscript $n$, such that
\begin{equation}
\label{eq:G_solutions}
\begin{split}
\psi_n(r,x) &= r J_1 (j_{1,i(n)} \frac{r}{r_0}) \sin \left( \left(j(n)-\frac{1}{2} \right) \pi x\right) \\ \lambda_{n} &= \left( \frac{j_{1,i(n)}}{r_0} \right)^2 + (2j(n)-1)^2 \frac{\pi^2}{4}\\
i(n) &= \text{floor}(n/N_r) + 1 \\ j(n) &= n - i(n) N_r + 1, \quad n \geq 0
\end{split}
\end{equation}\stepcounter{eqncounter}
The orthogonality relation for the eigenfunctions is
\begin{equation}
\label{eq:orthogonality}
\begin{split}
&<\psi_n(r,x),\psi_m(r,x)> = \\ &\int_0^{r_0} \int_0^1 \frac{1}{r}\psi_n(r,x) \psi_m(r,x) dr dx = \Lambda_{n}\delta_{nm},
\end{split}  
\end{equation}\stepcounter{eqncounter}
where $\Lambda_{n} = <\psi_n, \psi_n>$.

Now revisiting Eq. \ref{eq:axi_euler2}, after some reorganization it can be written as
\begin{equation}
\label{eq:euler3}
\begin{split}
& D_t G \psi_H +\\ & \frac{u_0(t)}{r} ( D_r \chi_0 D_x G - D_x \chi_0 H G + D_x G \chi_0 D_r - H G \chi_0 D_x ) \psi_H \\ 
& +\frac{1}{r}\left( D_r \psi_H D_x G \psi_H - D_x \psi_H H G \psi_H\right) \\
& +\frac{ u_0(t)^2}{r}\left( D_r \chi_0 D_x G \chi_0 - D_x \chi_0 H G \chi_0 \right) + \dot{u}_0(t) G\chi_0 = 0
\end{split}
\end{equation}\stepcounter{eqncounter}
Expanding the solutions onto the eigenfunctions of $G$,  
\begin{equation}
\label{eq:psi_expansion_appendix}
\psi_H(r,x,t)=\sum_{m} \eta_{m}(t)\psi_m(r,x)
\end{equation}\stepcounter{eqncounter}
and substituting this expression into \ref{eq:euler3},
\begin{equation}
\begin{split}
& \sum_{m} \lambda_{m}\dot{\eta}_{m}(t)\psi_m \\
& + \frac{u_0(t)}{r} \sum_{m} \eta_{m}(t) ( D_r \chi_0 D_x G - D_x \chi_0 H G ) \psi_m \\
& + \frac{u_0(t)}{r} \sum_{m} \eta_{m}(t) ( D_x G \chi_0 D_r - H G \chi_0 D_x ) \psi_m \\
& +\frac{1}{r} \sum_{m,l} \eta_{m}(t)\eta_{l}(t)( D_r \psi_m D_x G \psi_l - D_x \psi_m H G \psi_l) \\ 
& +  \frac{ u_0(t)^2}{r}\left( D_r \chi_0 D_x G \chi_0 - D_x \chi_0 H G \chi_0 \right) + \dot{u}_0(t) G \chi_0 =0
\end{split}
\end{equation}\stepcounter{eqncounter}
Taking the inner product of this equation with $\psi_n$ the time domain set of equations governing the evolution of $\eta_{n}(t)$ are obtained.
\begin{equation}
\label{eq:stream_modes_appendix}
\begin{split}
\dot{\eta}_{n}(t) + u_0(t) \sum_{m=1}^{N_{\eta}} B_{mn} \eta_{m}(t) + \sum_{l,m=1}^{N_{\eta}}
C_{lmn} \eta_{m}(t)\eta_{l}(t) + \\ u_0(t)^2 d_{n} + \dot{u}_0(t) f_{n}= 0
\end{split}
\end{equation}\stepcounter{eqncounter}
In this set of equations the constants encode the spatial information of the problem and are written as
\begin{equation}
\label{eq:eta_spatial_constants}
\begin{split}
\Lambda_{n} &= <\psi_n, \psi_n> \\
B_{mn} &= \frac{1}{\lambda_{n} \Lambda_{n}}<\psi_n,\frac{1}{r}( \lambda_{m} D_r \chi_0 D_x  - \lambda_{m} D_x \chi_0 H ) \psi_m> \\ & + \frac{1}{\lambda_{n} \Lambda_{n}}<\psi_n, \frac{1}{r}(D_x G \chi_0 D_r - H G \chi_0 D_x ) \psi_m> \\
 C_{lmn} &= \frac{1}{\lambda_{n} \Lambda_{n}} <\psi_n, \frac{\lambda_{l} }{r}(D_r \psi_m D_x \psi_l - D_x \psi_m H \psi_l)> \\
 d_{n} &= \frac{1}{\lambda_{n} \Lambda_{n}} <\psi_n, \frac{1}{r}(D_r \chi_0 D_x G \chi_0 - D_x \chi_0 H G \chi_0 )> \\
 f_{n} &= \frac{1}{\lambda_{n} \Lambda_{n}} <\psi_n, G \chi_0>
\end{split}
\end{equation}\stepcounter{eqncounter}

\subsection{The Vortical Pressure Source and Relation to the Acoustic Flow}
We are now in a position to express this pressure source in terms of $\boldsymbol{\eta}$ and $u_0$. In the previous section the stream function was expressed as
\begin{equation}
\psi(r,x,t) = \sum_n \eta_n(t) \psi_n (r,x) + u_0(t) \chi (r,x)
\end{equation}\stepcounter{eqncounter}
Inserting this expression into Eq. \ref{eq:p_src_appendix}, the pressure source can be expressed as 
\begin{equation}
\label{eq:p_src_constants}
\begin{split}
p_{src} &= -\sum_n \zeta_n \dot{\eta}_n - \dot{u}_0 \xi \\
\zeta_n &= \frac{1}{r_0^2}\int_0^{r_0}\int_0^1 r^2 \nabla^2 \psi_n(r,x) dr dx \\
\xi &= \frac{1}{r_0^2}\int_0^{r_0}\int_0^1 r^2 \nabla^2 \chi_0(r,x) dr dx
\end{split}
\end{equation}\stepcounter{eqncounter}
Substituting Eq. \ref{eq:stream_modes_appendix} for $\dot{\eta}_n$ into the expression for $p_{src}$,
\begin{dmath}
p_{src} = u_0 \sum_{m,n=1}^{N_{\eta}} B_{mn} \zeta_n \eta_{m} + \sum_{l,m,n=1}^{N_{\eta}}
C_{lmn} \zeta_n \eta_{m} \eta_{l} + u_0^2 \sum_{n=1}^{N_{\eta}} d_{n} \zeta_n + \dot{u}_0 \left( \xi + \sum_{n=1}^{N_{\eta}} f_{n} \zeta_n \right)
\end{dmath}\stepcounter{eqncounter}
Then substituting in Eq. \ref{eq:u_dot_appendix} for the $\dot{u}_0$ term,
\begin{dmath}
p_{src} = u_0 \sum_{m,n=1}^{N_{\eta}} B_{mn} \zeta_n \eta_{m} + \sum_{l,m,n=1}^{N_{\eta}} C_{lmn} \zeta_n \eta_{m} \eta_{l} + \\ u_0^2 \sum_{n=1}^{N_{\eta}} \left(d_{n} \zeta_n + \frac{\gamma}{2\mu} \left( \xi + \sum_{n=1}^{N_{\eta}} f_{n} \zeta_n \right) \right) + \\ \mu^{-1} \left(p - p_{src} + \sum_{i}^{N_q} \dot{q}_i \right) \left( \xi + \sum_{n=1}^{N_{\eta}} f_{n} \zeta_n \right).
\end{dmath}\stepcounter{eqncounter}
Solving this for $p_{src}$,
\begin{dmath}
\label{eq:p_src_appendix2}
p_{src} = c_1 u_0^2 + c_2 \left(p + \sum_{i}^{N_q} \dot{q}_i \right) + \\ c_3 \left(u_0 \sum_{m,n=1}^{N_{\eta}} B_{mn} \zeta_n \eta_{m} + \sum_{l,m,n=1}^{N_{\eta}} C_{lmn} \zeta_n \eta_{m} \eta_{l}  \right).  
\end{dmath}\stepcounter{eqncounter}
Here the constants are given by
\begin{equation}
\label{eq:c_constants}
\begin{split}
c_1 &= \frac{ \left(\textbf{d}+\frac{\gamma}{2\mu}\textbf{f}  \right) . \boldsymbol{\zeta} + \frac{\gamma}{2\mu} \xi }{1 + \frac{\xi + \textbf{f} . \boldsymbol{\zeta}}{\mu} } \\
c_2 &= \frac{\xi + \textbf{f} . \boldsymbol{\zeta}}{\mu+\xi + \textbf{f} . \boldsymbol{\zeta}} \\
c_3 &= \frac{1}{1 + \frac{\xi + \textbf{f} . \boldsymbol{\zeta}}{\mu}}
\end{split}
\end{equation}\stepcounter{eqncounter}
We will also need the time derivative of $p_{src}$,
\begin{dmath}
\label{eq:dot_p_src}
\dot{p}_{src} = 2 c_1 u_0 \dot{u}_0 + c_2 \sum_{i}^{N_q} \dot{q}_k + \\ c_3 \frac{\partial }{\partial t}\left(u_0 \sum_{m,n=1}^{N_{\eta}} B_{mn} \zeta_n \eta_{m} + \sum_{l,m,n=1}^{N_{\eta}} C_{lmn} \zeta_n \eta_{m} \eta_{l}  \right).  
\end{dmath}\stepcounter{eqncounter}

Now that we have derived a set of time domain equations, which govern the evolution of the vortical flow (described by $\boldsymbol{\eta}(t)$, we can express $p_{src}$ in terms of $u_0$, $\textbf{q}$, and $\boldsymbol{\eta}(t)$. We can also explicitly write out the dependence of the $\ddot{u}_0$ term on $\ddot{q}_m$ in Eq. \ref{eq:viscous_approximation_appendix}. This will allow us to algebraically solve Eq. \ref{eq:viscous_approximation_appendix} for $\ddot{\textbf{q}}$, which is necessary to input into a numerical solver, since the $\ddot{\textbf{q}}$ terms are dependent variables in a numerical solver.  

\begin{figure}
\centering
\includegraphics[width=\columnwidth]{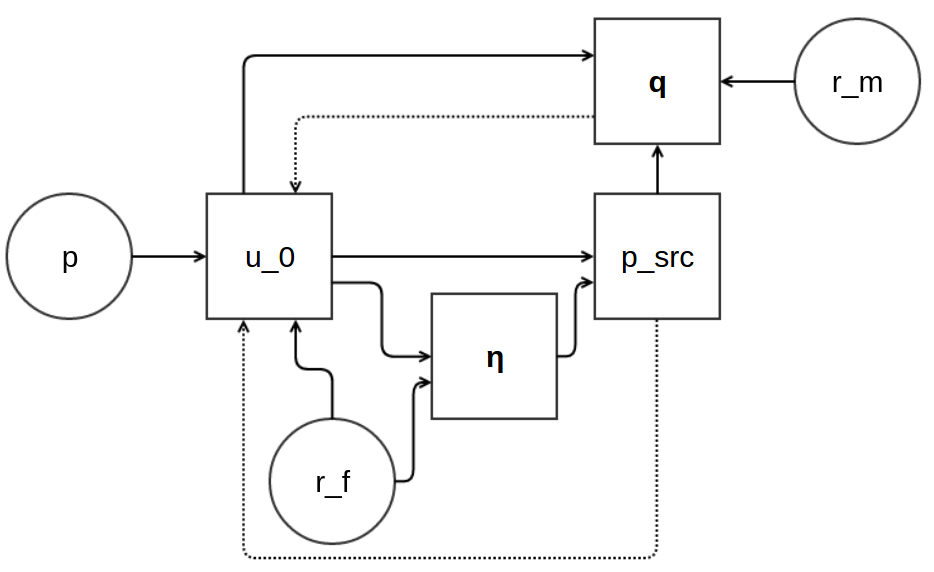}
\caption{Feedback diagram of the different variables describing the flow in the rodent vocal tract. Time dependent variables of the model are shown in boxes. Solid lines between the boxes indicate excitatory mechanisms, and dotted lines indicate inhibitory ones. The three input parameters of the model are shown inside the circles. These parameters are subglottal pressure $p$, vocal fold radius $r_F$, and mouth radius $r_m$. The subglottal pressure drives the velocity of the vocal fold flow $u_0$. The vocal fold flow then directly drives the vortical flow $\boldsymbol{\eta}$ and the acoustic flow $\textbf{q}$. The vocal fold radius modulates the strength of $u_0$ and $\boldsymbol{\eta}$. The evolution of $u_0$ and $\boldsymbol{\eta}$ determine the strength of $p_{src}$. The pressure $p_{src}$ then drives the acoustic flow $\textbf{q}$, which is modulated by the mouth radius, which controls the strength of radiation. The velocity $u_0$ is inhibited by $\textbf{q}$ and $p_{src}$ through pressure feedback.}
\label{fig:feedback}
\end{figure}

To begin fist differentiate $\dot{u}_0$ with respect to time,
\begin{equation}
\label{eq:u_ddot1}
\ddot{u}_0 = \mu^{-1} \left(-\dot{p}_{src} + \sum_{i}^{N_q} \dot{q}_k - \gamma u_{0} \dot{u}_0  \right).
\end{equation}\stepcounter{eqncounter}
Inserting Eq. \ref{eq:dot_p_src} for $\dot{p}_{src}$,
\begin{dmath}
\label{eq:u_ddot3}
\ddot{u}_0 = \frac{1-c_2}{\mu} \sum_{i}^{N_q} \dot{q}_k + \frac{(2 c_1 - 1) \gamma}{\mu} u_0 \dot{u}_0 + \frac{c_3 \gamma}{\mu} \frac{\partial }{\partial t}\left(u_0 \sum_{m,n=1}^{N_{\eta}} B_{mn} \zeta_n \eta_{m} + \sum_{l,m,n=1}^{N_{\eta}} C_{lmn} \zeta_n \eta_{m} \eta_{l}  \right).
\end{dmath}\stepcounter{eqncounter}
Now inserting this into Eq. \ref{eq:viscous} and gathering the $\ddot{q}_k$ terms together,
\begin{dmath}
M_{jk}^{-1}\ddot{q}_k + \beta_j \dot{q}_j +(\omega_j^2 - \alpha_j^2) q_j = -a_j' u_0 + b_j' F_b,
\end{dmath}\stepcounter{eqncounter} 
where
\begin{equation}
\begin{split}
\label{eq:Fb_appendix}
M_{jk}^{-1} &= \delta_{jk} + \frac{(c_2-1)b_j'}{\mu} \\
F_b &=  \frac{(2 c_1 - 1) \gamma}{\mu} u_0 \dot{u}_0 + \\ & \frac{c_3 \gamma}{\mu} \frac{\partial }{\partial t}\left(u_0 \sum_{m,n=1}^{N_{\eta}} B_{mn} \zeta_n \eta_{m} + \sum_{l,m,n=1}^{N_{\eta}} C_{lmn} \zeta_n \eta_{m} \eta_{l}  \right) 
\end{split}
\end{equation}\stepcounter{eqncounter}
Now multiply both sides of this equation by the matrix $\textbf{M}$ and writing the result in vector form.
\begin{equation}
\label{eq:acoustic_oscillator_appendix}
\boldsymbol{\ddot{q}} + \boldsymbol{D} \boldsymbol{\dot{q}} + \boldsymbol{K} \boldsymbol{q} = -\textbf{a} u_0 - \textbf{b} F_b,
\end{equation}\stepcounter{eqncounter}
where
\begin{equation}
\label{eq:acoustic_tensors}
\begin{split}
D_{ij} &= M_{ij} \beta_j \\
K_{ij} &= M_{ij} (\omega_j^2 - \alpha_j^2) \\
a_i &= \sum_m M_{ij} a_j' \\
b_i &= \sum_m M_{ij} b_j'
\end{split}
\end{equation}\stepcounter{eqncounter}
Eqs. \ref{eq:u_dot}, \ref{eq:stream_modes}, and \ref{eq:acoustic_oscillator}, with $p_{src}$ given by Eq. \ref{eq:p_src_appendix2} fully determine the dynamics of the rodent vocal tract. Fig. \ref{fig:feedback} is a feedback diagram of the variables and parameters present in the model. 
\stepcounter{figcounter}

\subsection{Analysis and Discussion}
\begin{figure}[!ht]
\centering
\includegraphics[width=\columnwidth]{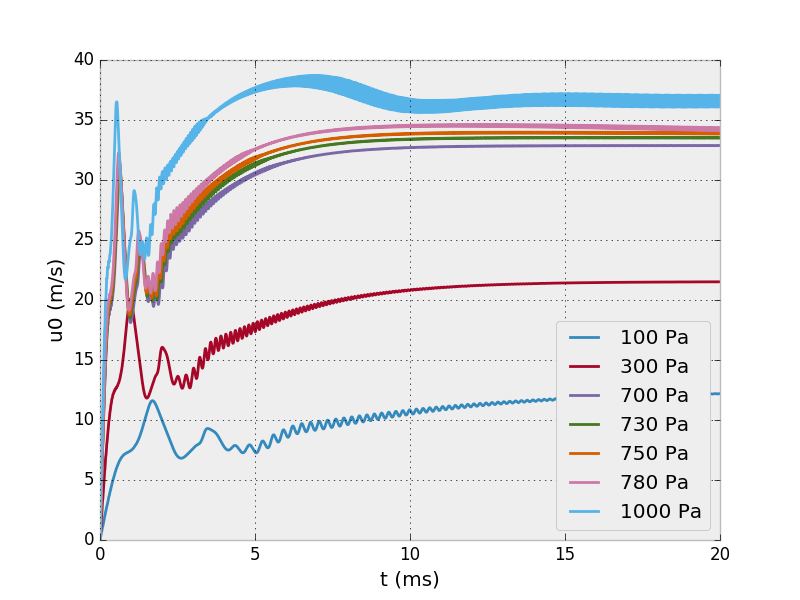}
\caption{The jet velocity entering the pharynx $u_0$ for different values of the subglottal pressure $p$. Somewhere around $710$ Pa the system transitions from approaching a fixed point to approaching a steady oscillatory state. In this calculation the vocal tract length is $L=4$ mm, the pharyngeal radius is $r_0=1.5$ mm, the tracheal radius is $r_t=1.5$ mm, the mouth radius is $r_{m}=.45$ mm, and the vocal fold radius is $r_f=1.1$ mm. }
\label{fig:pressure_sweep_u0}
\end{figure}

\begin{figure}[!ht]
\centering
\includegraphics[width=\columnwidth]{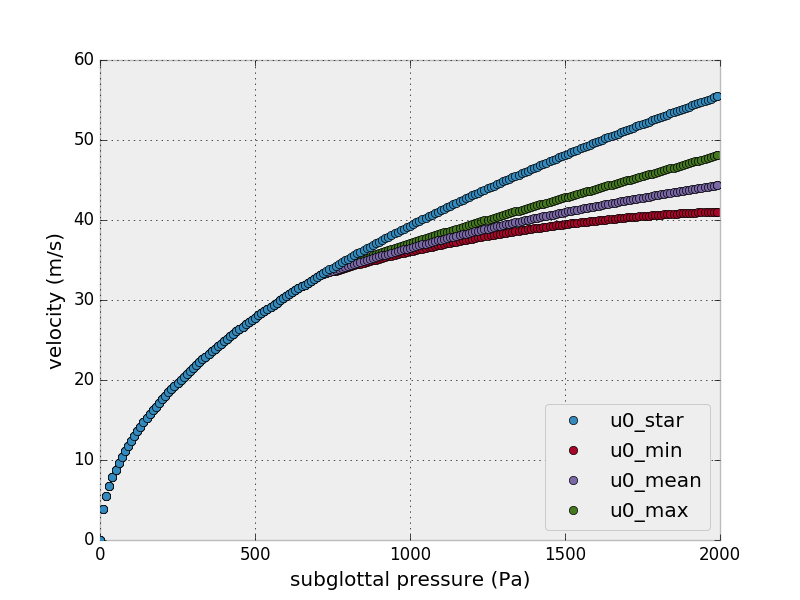}
\caption{The value of $u_0$ at the fixed point that system approaches when it is integrated from rest. Blue dots indicate the value of $u_0$ at the fixed point. Also shown are a minimum, mean, and max of the oscillations after the initial transients has died out. It can be seen that in its oscillating state the system does not orbit the unstable fixed point but rather some point nearby.}
\label{fig:fixed_point}
\end{figure}

Eqs. \ref{eq:u_dot}, \ref{eq:stream_modes}, and \ref{eq:acoustic_oscillator}, with $p_{src}$ given by Eq. \ref{eq:p_src_appendix2} can be integrated to determine the transient and steady state behavior of the system. Fig. \ref{fig:pressure_sweep_u0} shows the results of integrating the system from rest for several values of the subglottal pressure. For low values of $p$ the the solutions approach a fixed point in the steady state. For higher values of $p$ the steady state is a limit cycle. The limit cycle can be clearly seen at $p=1000$ Pa, but it is hard to calculate the threshold subglottal pressure this way.
\stepcounter{figcounter}
The onset of oscillations can more easily be seen by examining the behavior of the fixed point, which the system approaches when it is integrated from rest, and the behavior of the dominant eigenvalue of the Jacobian of the system, evaluated at that fixed point. Fig. \ref{fig:fixed_point} shows the value of $u_0$ at this fixed point. Also shown are the minimum, mean, and max of the oscillations after the initial transients has died out. The onset of oscillations can be seen where these quantities diverge from the fixed point. In its oscillating state the system does not orbit the unstable fixed point but rather some point nearby. This indicates that oscillations do not begin in a Hopf bifurcation but rather through some other process. They dynamics of this system will need to be explored in a later paper.
\stepcounter{figcounter}

\begin{figure}[!ht]
\centering
\includegraphics[width=\columnwidth]{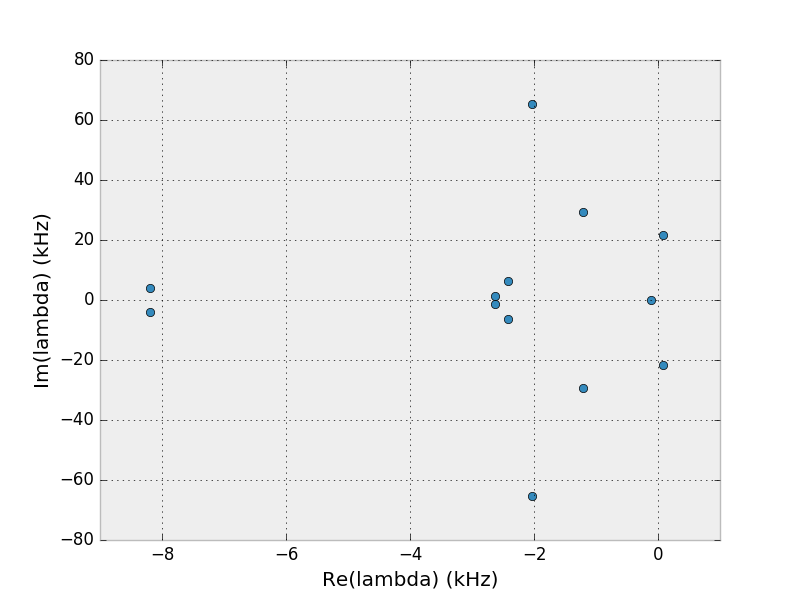}
\caption{The eigenvalue spectrum of the Jacobian evaluated at the fixed point for $p=1.5$ Pa. The system is driven by eigenvalues with the largest real part.}
\label{fig:ev_spectrum}
\end{figure}

The dominant eigenvalue $\lambda_d$ is the one with the largest real part and is associated with the vorticity oscillations caused by vortex ring formation. These oscillations drive the passive acoustic modes, all of which have eigenvalues with negative real part. Fig. \ref{fig:ev_spectrum} shows the full eigenvalue spectrum of the Jacobian evaluated at that fixed point for $p=1500$ Pa.  Fig. \ref{fig:stability} shows the real and imaginary parts of $\lambda_d$ as the subglottal pressure is varied. It can be seen that $Re(\lambda_d) < 0$ for low values of the subglottal pressure. The integration will approach a fixed point for these values of $p$. At around $p=710 Pa$ the real part of $\lambda_d$ crosses the imaginary axis, and the fixed point loses stability. From that point the system is driven at the frequency given by the imaginary part of $\lambda_d$. The subglottal pressure during rodent USVs has been measured to be about $p=1500$ Pa \cite{Riede2011}. It can be seen that as the subglottal pressure in the model approaches the biologically realistic value the driving frequency is approximately 22 kHz, which is exactly the frequency of the rat alarm call! It is instructive to compare the eigenvalue spectrum to the actual frequency spectrum of $u_0$. Fig. \ref{fig:specgram} shows the steady state frequency spectrum of $u_0$. The dominant eigenvalues interact with the passive acoustic ones to produce the harmonic content.
\stepcounter{figcounter}

\end{document}